\documentstyle[12pt]{article}

\newcommand{\be}{\begin{equation}}
\newcommand{\ee}{\end{equation}}
\newcommand{\ba}{\begin{eqnarray}}
\newcommand{\ea}{\end{eqnarray}}
\begin{document}
\begin{center}
{\bf DYNAMICS OF COSMOLOGICAL PHASE TRANSITIONS}\footnote[1]{Based on
Lectures Delivered at the XIII Course of the International School of
 Cosmology and Gravitation,
Erice, May 1993, and at the 17th Johns Hopkins Workshop on Current
Problems in Particle Theory, July 1993.}
\end{center}
\begin{center}
{{\bf D. Boyanovsky$^{(a)}$,  H. J. de Vega$^{(b)}$ and R.
 Holman$^{(c)}$}}
\end{center}
\begin{center}
{\it (a) Department of Physics and Astronomy, University of
Pittsburgh\\Pittsburgh, P. A. 15260, U.S.A.} \\
{\it (b) Laboratoire de Physique Theorique et Hautes
Energies\cite{lpthe}\\Universite
Pierre et Marie Curie\\Tour 16, 1er etage, 4, Place
Jussieu\\75252 Paris Cedex 05, France.}\\
{\it (c) Department of Physics, Carnegie Mellon University\\
Pittsburgh, P.A. 15213, U.S.A}
\end{center}
\begin{abstract}
The dynamics of typical phase transitions is studied out of
 equilibrium in
weakly coupled inflaton-type scalar field theories in Minkowski
space. The shortcomings of the effective potential and equilibrium
descriptions are pointed out. A case of a rapid supercooling from
$T_i>T_c$ to $T_f \ll T_c$ is considered.
 The equations of motion up to one-loop for the
order parameter are obtained and integrated for the case of ``slow
rollover initial conditions''. It is shown that the instabilities
responsible for the process of phase separation introduce dramatic
corrections to the evolution. Domain formation and
growth (spinodal decomposition) is studied in a
 non-perturbative self-consistent approximation.
 For very weakly coupled theories domains grow for a long time, their
final size is several times the zero temperature correlation length. For
strongly coupled theories the final size of the domains is comparable to
the zero temperature correlation length and the transition proceeds faster.
We also obtain the evolution equations for the order parameter and the
fluctuations to one-loop order and in a
non-perturbative Hartree approximation in spatially flat FRW cosmologies.
The renormalization, and leading behavior of the high temperature limit
are analyzed.
\end{abstract}

\section{\bf Introduction and Motivation}

Inflationary cosmological models provide very appealing
scenarios
to describe the early evolution of our
universe\cite{abbott}.
Since the original model proposed by Guth\cite{guth1},
several
alternative scenarios have been proposed to overcome some of
the
difficulties with the original proposal.

Among them, the new inflationary
model\cite{guth2,linde1,steinhardt,linde2}
is perhaps one of the most attractive. The essential
ingredient
in the new inflationary model is a scalar field (the ``inflaton'') that
undergoes a
second order phase transition from a high temperature
symmetric phase
to a low temperature broken symmetry phase. The expectation
value (or
thermal average) of the scalar field $\phi$ serves as the
order parameter.
Initially at high temperatures, the scalar field is assumed
to be in
thermal equilibrium and  $\phi \approx 0$.  The usual field-
theoretic tool
to study the phase transition is the effective
potential\cite{kirzhnits,dolan,sweinberg}.

 At high
temperatures,
the global minimum of the effective
potential is at $\phi =0$, whereas at low temperatures there
are two
degenerate minima.

The conjectured behavior of the phase transition in the new inflationary
model is the
following: as the universe cools down, the expectation value
of the scalar
field
remains close to zero until the temperature becomes smaller
than the
critical temperature at which
the effective
potential develops degenerate minima away from the origin.
When this happens, the scalar
field begins to ``roll down the potential hill''. In the new
inflationary
scenario, the effective potential below the critical
temperature is
extremely flat near the maximum, and the scalar field
remains near the
origin, i.e. the false vacuum, for a very long
time and eventually rolls down the hill very slowly, hence the name
 ``slow rollover''.
During this stage, the energy density of
the universe is dominated by the constant  energy
density of the false vacuum $V_{eff}(
\phi=0)$, and the universe evolves rapidly into a de Sitter
space (see for
example the reviews by Kolb and Turner\cite{kolb}, Linde\cite{linde}
and Brandenberger\cite{brandenberger}). Perhaps the most remarkable
consequence of the new
inflationary scenario and the slow rollover transition is
that they provide a calculational framework for the prediction
of density fluctuations\cite{starobinsky}. The coupling
constant in the
typical zero temperature potentials must be fine tuned to a
very small value to reproduce the observed limits on  density
fluctuations\cite{kolb,linde}.

The slow rollover scenario is based on the
{\it static}
effective potential. Its  usefulness in a time dependent situation
 has been questioned by Mazenko,
Unruh and
Wald\cite{mazenko} who argued that the {\it
dynamics} of the
cooling down process is very similar to the process of phase
separation in
statistical mechanics, and  that the system will form
domains within which the scalar field will relax to the values
at the minima of
the potential very quickly. It is now accepted that this picture
may be correct for very strongly coupled theories but not for
weakly coupled scenarios as required in inflation.

Guth and Pi\cite{guthpi}, studied
the effects
of quantum fluctuations on the time evolution below the critical
 temperature by treating the
potential
near the origin as an {\it inverted harmonic oscillator}.
They recognized that the
instabilities associated with these upside-down oscillators
lead to an
exponential growth of the quantum fluctuations at long times
and to a
classical description of the probability distribution
function. Guth
and Pi also recognized that the {\it static} effective
potential is not
appropriate to describe the dynamics, that must be treated
as a time
dependent process.

Subsequently, Weinberg and Wu\cite{weinbergwu}, have studied
the effective
potential, particularly in the situation when the tree level
potential
allows for broken symmetry ground states. In this case, the effective
potential becomes complex. These authors identified the imaginary
part of the effective potential with the decay rate of a particular
initial state.

In statistical mechanics, the imaginary part is a consequense of a
sequence of thermodynamically unstable states.

The effective potential offers no information on the
{\it dynamics} of the process of the phase transition.
As mentioned above it becomes complex in a region in field space
corresponding to thermodynamically unstable states.

Furthermore, in an expanding gravitational background the notion of
a static effective potential is at best an approximate one.
To understand whether one may treat the problem in the approximation
of local thermodynamic equilibrium, the time scales involved must
be understood carefully. In a typical FRW cosmology the important time
scale is determined by the Hubble expansion factor $H(t) = \dot{a}(t) /
a(t)$ whereas the equilibration processes are determined by the
typical collisional relaxation rates $\Gamma(T) \approx \lambda^2 T$
in typical scalar theories. Local thermodynamic equilibrium
will prevail if $\Gamma(T) \gg H(t)$. In de Sitter evolution, after a
typical phase transition at $T \approx T_c \approx 10^{14}$ Gev, $H
\approx 10^{-5} T_c$\cite{linde} and local thermodynamic equilibrium
{\it will not} prevail for weakly coupled inflaton theories in which
$\lambda \approx 10^{-12}-10^{-14}$\cite{linde,starobinsky,weakcoup}.

Thus in these scenarios, weakly coupled scalar field theories {\it
cannot be assumed} to be in local thermal equilibrium. Typical expansion
rates are much larger than typical equilibration rates and the
phase transition will occur very rapidly. The
 long-wavelength fluctuations will be strongly out of equilibrium as
they typically have very slow dynamics (see\cite{boyshin} and references
therein). The phase transition {\it must} be studied away from
 equilibrium.

A complete discussion of these issues and the complex effective potential
may be found in the articles by Boyanovsky and de Vega\cite{boyveg}
and Boyanovsky et. al.\cite{boyshin,boya}.

\section{\bf Non-Equilibrium time evolution}

 Let us consider
the situation in which for time $t<0$ the system is in {\it
equilibrium} at an initial temperature $T_i > T_c$ where
$T_c$ is the critical temperature.
At time $t = 0$ the system is  rapidly ``quenched''
to a final temperature below the critical temperature
$T_f < T_c$ and evolves thereafter out of equilibrium.

What we have in mind in this situation, is a cosmological
scenario with a period
of rapid inflation in which the temperature drops very fast
compared to typical relaxation times of the scalar field.

 Precisely because of the weak couplings and critical
slowing down of long-wavelength fluctuations,
we conjecture that an inflationary period at
temperatures near
the critical temperature, may be described in this
``quenched'' approximation.
Another situation that may be described by this
approximation is
that of a scalar field again at $T_i>T_c$ suddenly coupled
 to a ``heat bath'' at a
much lower temperature (below the transition temperature)
and evolving out of equilibrium.

 The ``heat bath'' could consist of other
fields at a different temperature.

To understand whether this quenched approximation is valid or not
will require a deeper
understanding of the initial conditions.

Although we are currently studying the case of inflationary
cosmologies, we will here concentrate on the dynamics of a supercooled
phase transition in Minkowski space.

This situation  is modelled by introducing a Hamiltonian with a
{\it time dependent mass term}
\begin{eqnarray}
  H(t) & = & \int_{\Omega} d^3x \left\{
\frac{1}{2}\Pi^2(x)+\frac{1}{2}(\vec{\nabla}\Phi(x))^2+\frac
{1}{2}m^2(t)
\Phi^2(x)+\frac{\lambda}{4!}\Phi^4(x) \right \}
\label{timedepham} \\
m^2(t) & = & m^2 \Theta(-t) + (-\mu^2) \Theta(t)
\label{massoft}
\end{eqnarray}
where both $m^2$ and $\mu^2$ are positive. We assume that
for all times $t <0$ there is thermal equilibrium at
temperature $T_i$, and the system is described by the
density matrix
\begin{eqnarray}
\hat{\rho}_i  & = & e^{-\beta_i H_i} \label{initialdesmat}\\
          H_i & = & H(t<0) \label{initialham}
\end{eqnarray}
In the Schroedinger picture, the density matrix evolves in
time as
\begin{equation}
\hat{\rho(t)} = U(t)\hat{\rho}_iU^{-1}(t) \label{timedesmat}
\end{equation}
with $U(t)$ the time evolution operator.

An alternative and equally valid interpretation (and the one
that
we like best) is that the initial
condition  being considered here is that of a system in
equilibrium in the
symmetric phase which is then evolved in time with a Hamiltonian that
allows for
broken symmetry ground states, i.e.  the Hamiltonian (\ref
{timedepham}, \ref{massoft}) for $t>0$.

The expectation
value of any operator is thus
\begin{equation}
< {\cal{O}} >(t) = Tr e^{-\beta_i H_i} U^{-1}(t)
{\cal{O}}U(t)/ Tr e^{-
\beta_i H_i}
\label{expecvalue}
\end{equation}
This expression may be written in a more illuminating form
by
choosing an arbitrary time $T <0$ for which
$U(T) = \exp[-iTH_i]$ then we may write
$\exp[-\beta_i H_i] = \exp[-iH_i(T-i\beta_i -T)] = U(T-
i\beta_i,T)$.
Inserting in the trace $U^{-1}(T)U(T)=1$,
commuting $U^{-1}(T)$ with $\hat{\rho}_i$ and using the
composition property of the evolution operator, we may write
(\ref{expecvalue}) as
\begin{equation}
< {\cal{O}}>(t) = Tr U(T-i\beta_i,t) {\cal{O}} U(t,T)/ Tr
U(T-i\beta_i,T) \label{trace}
\end{equation}
The numerator of the above expression has a simple meaning:
start at time $T<0$, evolve to time $t$, insert the operator
$\cal{O}$ and evolve backwards in time from $t$ to $T<0$,
and along the negative imaginary axis from $T$ to $T-
i\beta_i$.
The denominator, just evolves along the
negative imaginary axis from $T$ to $T-i\beta_i$. The
contour in
the numerator may be extended to an arbitrary large positive
time
$T'$ by inserting $U(t,T')U(T',t)=1$ to the left of
$\cal{O}$ in
(\ref{trace}), thus becoming
\begin{equation}
< {\cal{O}}>(t) = Tr U(T-
i\beta_i,T)U(T,T')U(T',t){\cal{O}}U(t,T)
/Tr U(T-i\beta_i,T)
\end{equation}
The numerator now represents the process of evolving from
$T<0$
to $t$, inserting the operator $\cal{O}$, evolving further
to
$T'$, and backwards from $T'$ to $T$ and down the negative
imaginary axis to $T-i\beta_i$. Eventually we take $T \rightarrow -
\infty
\; ; \; T' \rightarrow \infty$. It is straightforward to
generalize
to  real time correlation functions of Heisenberg picture
operators.

As usual, the insertion of an operator may be achieved by
inserting sources in the time evolution operators, defining
the generating functionals and eventually taking functional
derivatives with respect to these sources. Notice that we
have
three evolution operators, from $T$ to $T'$, from $T'$, back
to $T$ (inverse operator) and from $T$ to $T-i\beta_i$.

Since
each of these operators has interactions and we want  to
 generate the diagrammatics of perturbation theory from the
generating functionals, we use {\it three different
sources}.
A source $J^+$
for the evolution  $T \rightarrow T'$, $J^-$ for the branch
$T'\rightarrow T$ and finally $J^{\beta}$ for
$T \rightarrow T- i\beta_i$. The denominator may be obtained
from the numerator by
setting $J^+ = J^- =0$. Finally the generating functional
$Z[J^+, J^-, J^{\beta}]= Tr U(T-i\beta_i,T;J^{\beta})
U(T,T';J^-)U(T',T;J^+)$, may be written in term of path
integrals as (here we neglect the spatial arguments to avoid
cluttering of notation)
\begin{eqnarray}
Z[J^+,J^-,J^{\beta}] & = & \int D \Phi D \Phi_1 D \Phi_2
\int
{\cal{D}}\Phi^+ {\cal{D}}\Phi^-
{\cal{D}}\Phi^{\beta}e^{i\int_T^{T'}\left\{{\cal{L}}[\Phi^+,
J^+]-
{\cal{L}}[\Phi^-,J^-]\right\}}\times   \nonumber\\
                     &   &  e^{i\int_T^{T-
i\beta_i}{\cal{L}}[\Phi^{\beta}, J^{\beta}]}
\label{generfunc}
\end{eqnarray}
with the boundary conditions $\Phi^+(T)=\Phi^{\beta}(T-
i\beta_i)=\Phi \;
; \; \Phi^+(T')=\Phi^-(T')=\Phi_2 \; ; \; \Phi^-
(T)=\Phi^{\beta}(T)=
\Phi_1$.
As usual the path integrals over the quadratic forms may be
done and one obtains the final result for the partition
function
\begin{eqnarray}
Z[J^+,J^-, J^{\beta}] & = & e^{\left\{i\int_{T}^{T'}dt
\left[{\cal{L}}_{int}(-i\delta/\delta J^+)-
{\cal{L}}_{int}(i\delta/\delta
J^-)\right] \right \}}e^{\left\{i\int_{T}^{T-
i\beta_i}dt{\cal{L}}_{int}(-i\delta/\delta J^{\beta})
\right\}}
\times \nonumber \\
                      &   & e^{\left\{\frac{i}{2}\int_c
dt_1\int_c dt_2 J_c(t_1)J_c(t_2)G_c (t_1,t_2) \right\}}
 \label{partitionfunction}
\end{eqnarray}
Here $J_c $ are the currents defined on the segments of the
 contour\cite{boyveg}
 $J^{\pm}\; ,\; J^\beta$\cite{niemisemenoff}
and $G_c$ is the Green's function on the
contour (see below), and again the spatial arguments have
been suppressed.

In the two contour integrals (on $t_1 ; \; \; t_2$) in
(\ref{partitionfunction}) there are altogether nine terms,
corresponding to the combination of currents in each of the
three branches.
However, in the limit $T \rightarrow -\infty$, the
contributions
arising from the terms in which one current is on the $(+)$
or $(-)$ branch and another on the imaginary time segment
(from T to $T-i\beta_i$), go
to zero when computing correlation functions in which the
external
legs are at finite {\it real time}. For theses {\it real time
correlation functions} there is no contribution from the
$J^{\beta}$ terms. These cancel between numerator and
denominator, and the information on finite
temperature is encoded in the boundary conditions on the
Green's
functions (see below). Then for the calculation of finite
{\it real time} correlation functions the generating
functional
simplifies to\cite{calzetta,calzettahu}
\begin{eqnarray}
Z[J^+,J^-] & = & e^{\left\{i\int_{T}^{T'}dt\left[
{\cal{L}}_{int}(-i\delta/\delta J^+)-
{\cal{L}}_{int}(i\delta/\delta
J^-)\right] \right \}} \times \nonumber \\
           &   & e^{\left\{\frac{i}{2}\int_T^{T'}
dt_1\int_T^{T'} dt_2 J_a(t_1)J_b(t_2)G_{ab} (t_1,t_2)
\right\}}
 \label{generatingfunction}
\end{eqnarray}
with $a,b = +,-$.

This formulation in terms of time evolution along a contour
in complex time has been used many times in non-equilibrium
statistical mechanics. To our knowledge the first to use
this formulation were Schwinger\cite{schwinger} and
Keldysh\cite{keldysh} (see also Mills\cite{mills}).
 There are many articles in the literature
using these techniques to study time dependent problems,
some of the more clear articles are by Jordan\cite{jordan},
Niemi and Semenoff\cite{niemisemenoff},
Landsman and van Weert\cite{landsman},Semenoff and
Weiss\cite{semenoffweiss}, Kobes and
Kowalski\cite{kobeskowalski}, Calzetta and
Hu\cite{calzettahu},Paz\cite{paz} and references therein (for more
details see\cite{boyveg,boyshin,boya}.

  The Green's functions that enter in the integrals along
the  contours in (\ref{partitionfunction},
\ref{generatingfunction})
  are given by (see above references)
\begin{eqnarray}
G^{++}(t_1,t_2)  & = & G^{>}(t_1,t_2)\Theta(t_1 - t_2) +
G^{<}(t_1,t_2)\Theta(t_2-t_1) \label{timeordered}\\
G^{--}(t_1,t_2)  & = & G^{>}(t_1,t_2)\Theta(t_2-t_1) +
G^{<}(t_1,t_2)\Theta(t_1-t_2) \label{antitimeordered} \\
G^{+-}(t_1,t_2)  & = & -G^{<}(t_1,t_2) \label{plusminus}\\
G^{-+}(t_1,t_2)  & = & -G^{>}(t_1,t_2) = -G^{<}(t_2,t_1)
\label{minusplus}\\
G^{<}(T,t_2)     & = & G^{>}(T-i\beta_i,t_2)
\label{periodicity}
\end{eqnarray}

As usual $G^{<},G^{>}$ are homogeneous solutions of the
quadratic form with appropriate boundary conditions. We will
construct them explicitly later.
The condition (\ref{periodicity}) is recognized as the
periodicity condition in imaginary time (KMS
condition)\cite{kadanoffbaym}.

 To obtain the evolution equations we use the tadpole
 method\cite{sweinberg},
and write
\begin{equation}
\Phi^{\pm}(\vec{x},t) = \phi(t) + \Psi^{\pm}(\vec{x},t)
\label{expecvaluepsi}
\end{equation}
Where, again, the $\pm$ refer to the branches for forward
and backward time propagation. The reason for shifting both
($\pm$) fields by the {\it same} classical configuration, is
that $\phi$ enters in the time evolution operator as a
background c-number variable, and time evolution forward and
backwards are now considered in this background.

The  evolution equations are obtained with the
tadpole method by expanding the Lagrangian around $\phi(t)$
and considering the {\it linear}, cubic, quartic, and higher
order terms in $\Psi^{\pm}$ as perturbations and requiring
that
\[ <\Psi^{\pm}(\vec{x},t) >= 0 .\]

It is a straightforward exercise to see that this is
equivalent to
extremizing
the one-loop effective action in which the determinant (in
the logdet)
incorporates the boundary condition of equilibrium at time
$t<0$ at
the initial temperature.

To one loop we find the  equation of motion
\begin{equation}
\frac{d^2\phi(t)}{dt^2}+m^2(t)\phi(t)+\frac{\lambda}{6}
\phi^3(t)+ \frac{\lambda}{2}\phi(t)\int \frac{d^3
k}{(2\pi)^3}(-i) G_k(t,t) = 0 \label{eqofmotion}
\end{equation}
with $G_k(t,t)=G_k^{<}(t,t)=G_k^{>}(t,t)$ is the spatial
Fourier transform of the equal-time Green's function.

Notice that

\[(-iG_k(t,t)) =<\Psi^+_{\vec{k}}(t) \Psi^+_{-
\vec{k}}(t)>\]

is a {\it positive definite quantity} (because the field
$\Psi$ is real) and as we argued before
(and will be seen explicitly shortly) this Green's function
grows in time because of the instabilities associated with
the
phase transition and domain growth\cite{guthpi,weinbergwu}.

These Green's functions are constructed out of the
homogeneous solutions to the operator of quadratic
fluctuations
\begin{eqnarray}
\left[\frac{d^2}{dt^2} + \vec{k}^2 +
M^2(t)\right]{\cal{U}}_k^{\pm} & = & 0 \label{homogeneous}\\
 M^2(t)  =  (m^2+\frac{\lambda}{2}
\phi^2_i)\Theta(-t)            & + &
(-\mu^2+\frac{\lambda}{2}\phi^2(t))\Theta(t)
\label{bigmassoft}
\end{eqnarray}

The boundary conditions on the homogeneous solutions are
\begin{eqnarray}
{\cal{U}}_k^{\pm}(t<0) & = & e^{\mp i
\omega_{<}(k)t}\label{boundaryconditions} \\
\omega_{<}(k)          & = &
\left[\vec{k}^2+m^2+\frac{\lambda}{2}\phi^2_i\right]^{\frac{
1}{2}}\label{omegaminus}
\end{eqnarray}
where $\phi_i$ is the value of the classical field at time
$t<0$ and is the initial boundary condition on the equation
of motion. Truly speaking, starting in a fully symmetric
phase
will force $\phi_i =0$, and the time evolution will maintain
this value. Therefore we admit a small explicit symmetry
breaking field in the initial density matrix to allow for a
small $\phi_i$. The introduction of this initial condition
seems artificial since we are studying the situation of
cooling down from the symmetric phase. However,
 we recognize that  the phase transition from the
symmetric phase
 occurs via formation of domains (in the case of a discrete
symmetry)
inside which the order parameter acquires non-zero values.
The domains
will have the same probability for either value of the field
and the
volume average of the field will remain zero. These domains
will grow in time, this is the phenomenon of phase
separation and spinodal decomposition
familiar in condensed matter physics. Our evolution
equations presumably
will apply to the coarse grained average of the scalar field
inside each
of these domains. This average will only depend on time.
Thus, we interpret
$\varphi_i$ as corresponding to the coarse grained average
of the field in each of these domains.
The question of initial conditions on the scalar field is
also present (but usually overlooked) in the slow-rollover
scenarios but as we will see later, it plays a fundamental
role in the description of the evolution.

The identification of the initial value $\varphi_i$ with the
average of the field in each domain is certainly a
plausibility argument to
justify an initially small asymmetry in the scalar field
which is necesary
for the further evolution of the field, and is consistent
with the usual
assumption within the slow rollover scenario.

 The boundary conditions on the mode functions
${\cal{U}}_k^{\pm}(t)$ correspond to  ``vacuum'' boundary
conditions of positive and negative frequency modes
(particles and antiparticles) for $t<0$.

Finite temperature enters through the periodicity conditions
(\ref{periodicity}) and the Green's functions are
\begin{eqnarray}
G^{>}_k(t,t') & = & \frac{i}{2\omega_<(k)} \frac{1}{1-e^{-
\beta_i \omega_<(k)}}\left[{\cal{U}}_k^+(t) {\cal{U}}_k^-
(t')+ e^{-\beta_i\omega_<(k)} {\cal{U}}_k^-(t)
{\cal{U}}_k^+(t') \right] \label{finalgreenfunc} \\
G^{<}_k(t,t') & = & G^{>}(t',t)
\end{eqnarray}

The effective equations of motion to one loop
that determine the time evolution of the scalar field are
\begin{eqnarray}
\frac{d^2\phi(t)}{dt^2}+m^2(t)\phi(t) & + &
\frac{\lambda}{6}
\phi^3(t) + \nonumber \\
                                      &   &
               \frac{\lambda}{2}\phi(t) \int \frac{d^3
k}{(2\pi)^3} \frac{{\cal{U}}_k^+(t) {\cal{U}}_k^-(t)
}{2\omega_<(k)} \mbox{coth}\left[\frac{\beta_i
\omega_<(k)}{2}\right]
 =  0 \label{finaleqofmotion1} \\
\left[\frac{d^2}{dt^2} + \vec{k}^2 +
M^2(t)\right]{\cal{U}}_k^{\pm}
          & = & 0 \label{finaleqofmotion2}
\end{eqnarray}
with (\ref{bigmassoft}) , (\ref{boundaryconditions}).

This set of equations is too complicated to attempt an
analytic
solution. They will be dealt with numerically. However,
 before doing this, we should note that
there are several features of this set of equations
that reveal the basic physical aspects of the dynamics of
the
scalar field.

{\bf i)}: The effective evolution equations are {\bf real}.
The mode functions ${\cal{U}}_k^{\pm}(t)$ are complex
conjugates of  each other
as may be seen from the time reversal symmetry of the
equations,
and the boundary conditions (\ref{boundaryconditions}).
This  situation must be contrasted with the expression for
the
effective potential for the {\it analytically continued
modes}.

{\bf ii)}: Consider the situation in which the initial
configuration of the classical field is near the origin
$\phi_i \approx 0$,
for $t>0$. The modes for which $\vec{k}^2 < (k_{max})^2 \; ;
\;
(k_{max})^2 = \mu^2-\frac{\lambda}{2}\phi_i^2$ are {\it
unstable}.

In particular, for early times ($t>0$), when $\phi_i \approx
0$,
these unstable modes behave approximately as
\begin{eqnarray}
{\cal{U}}_k^+(t) & = & A_k e^{W_k t}+B_k e^{-W_k t}
\label{unstable1}\\
{\cal{U}}_k^-(t) & = & ({\cal{U}}_k^+(t))^{*}
\label{unstable2}\\
A_k              & = & \frac{1}{2}\left[1-
i\frac{\omega_{<}(k)}
{W_k} \right] \; ; \; B_k = A_k^* \label{abofk} \\
W_k              & = & \left[\mu^2-\frac{\lambda}{2}\phi_i^2
- \vec{k}^2 \right]^{\frac{1}{2}} \label{bigomega}
\end{eqnarray}
Then the early time behavior of $(-iG_k(t,t))$ is given by
\begin{equation}
(-iG_k(t,t)) \approx \frac{1}{2\omega_<(k)}
\left[1+\frac{\mu^2+m^2}
{\mu^2-\frac{\lambda}{2} \phi_i^2-k^2}[cosh(2W_kt)-1]\right]
coth[\beta_i\omega_<(k)/2]
\label{earlytimegreen}
\end{equation}
This early time behavior coincides with the Green's function
of Guth and
Pi\cite{guthpi} and Weinberg and Wu\cite{weinbergwu} for the
inverted harmonic oscillators when our initial
state (density matrix) is taken into account.

Our evolution equations, however, permit us to go beyond the
early
time behavior and to incorporate the non-linearities that
will eventually shut off the instabilities.

These early-stage instabilities and subsequent growth of
fluctuations
and correlations, are the hallmark of the process of phase
separation,
and precisely the instabilities that trigger the phase
transition.

It is
clear from the above equations of evolution, that the
description in terms
of inverted oscillators will only be valid at very early
times.
At such times, the {\it stable} modes for which
$\vec{k}^2>(k_{max})^2$
are obtained
from (\ref{unstable1}) , (\ref{unstable2}) , (\ref{abofk})
by the  analytic continuation
$ W_k \rightarrow -i \omega_{>}(k)= \left[ \vec{k}^2-\mu^2+
\frac{\lambda}{2}\phi_i^2 \right]^{\frac{1}{2}} $.

For $t<0$, ${\cal{U}}_k^+(t){\cal{U}}_k^-(t)=1$ and one
obtains
the usual result for the evolution equation
\[ \frac{d^2 \phi(t)}{dt^2}+\frac{dV_{eff}(\phi)}{d\phi} =0
\]
with $V_{eff}(\phi)$ the finite temperature effective
potential
but for $t<0$ there are no unstable modes.

It becomes clear, however, that for $t>0$ there are no {\it
static}
solutions to the evolution equations for $\phi(t) \neq 0$.

{\bf iii)}{\bf Coarsening}: as the classical expectation
value
$\phi(t)$ ``rolls down'' the potential hill, $\phi^2(t)$
increases and $(k_{max}(t))^2= \mu^2-
\frac{\lambda}{2}\phi^2(t)$
{\it decreases}, and only the very
long-wavelength modes remain unstable, until for a
particular
time  $t_s\; ; \; (k_{max}(t_s))^2=0$. This occurs when
$\phi^2(t_s) = 2\mu^2/\lambda$, which is the inflexion point
of the
tree level potential. In statistical mechanics this point is
known
as the ``classical spinodal point'' and $t_s$ as the
``spinodal time''\cite{langer,guntonmiguel}. When the
classical
field reaches the spinodal point, all instabilities shut-
off.
{}From this point on, the dynamics is oscillatory and this
period is
identified with the ``reheating'' stage in cosmological
scenarios\cite{linde,brandenberger}.

It is
clear from the above equations of evolution, that the
description in terms
of inverted oscillators will only be valid at small positive
times, as eventually
the unstable growth will shut-off.

The value of the spinodal time depends on the
initial conditions of $\phi(t)$. If the initial value
$\phi_i$ is
very near the classical spinodal point, $t_s$ will be
relatively
small and there will not be enough time for the unstable
modes
to grow too much.
In this case, the one-loop corrections for small coupling
constant
will remain perturbatively small.
On the other hand, however, if $\phi_i \approx 0$, and the
initial velocity is small, it will take a very long time to
reach the
classical spinodal point. In this case the unstable modes
{\bf may grow
dramatically making the one-loop corrections non-negligible
even for
small coupling}. These initial conditions of small initial
field and
velocity are precisely the ``slow rollover'' conditions that
are of interest in cosmological scenarios of ``new
inflation''.

The renormalization aspects have been studied in
reference\cite{boyveg} and we refer the reader to that article for
details.

 \section{\bf Analysis of the Evolution}

 As mentioned previously within the context of coarsening,
when
 the initial value of the scalar field $\phi_i \approx 0$,
and
 the initial temporal derivative is small, the scalar field
slowly
 rolls down the potential hill. But during the time while
the scalar
 field remains smaller than the ``spinodal'' value, the
unstable
 modes grow and the one-loop contribution grows
as a consequence. For a
 ``slow rollover'' condition, the field remains very small
($\phi^2(t)
 \ll 2\mu^2/\lambda$) for a long
 time, and during this time the unstable modes grow
exponentially.
After renormalization, the stable modes give an oscillatory
contribution
 which is bound in time, and for weak coupling remains
perturbatively
 small at all times.

 Then for a ``slow rollover'' situation and for
 weak coupling, only the unstable modes will yield to an
important
 contribution to the one-loop correction. Thus, in the
evolution equation
 for the scalar field, we will keep only the integral over
the {\it
 unstable modes} in the one loop correction.

Phenomenologically the coupling constant in these models is
bound by the spectrum of density fluctuations to be within
the range $\lambda_R \approx 10^{-12} - 10^{-
14}$\cite{linde}. The  stable modes will
\underline{always} give a \underline{perturbative}
contribution, whereas the unstable modes grow exponentially
in time thus raising the possibility of giving a non-
negligible contribution.

 With the purpose of numerical analysis of the effective
equations
 of motion, it proves convenient to introduce the following
 dimensionless variables
 \begin{eqnarray}
 \tau      & = & \mu_R t \; ; \;  q   =  k/\mu_R \\
 \eta^2(t) & = & \frac{\lambda_R}{6\mu_R^2}\phi^2(t) \; ; \;
L^2 =
 \frac{m_R^2 + \frac{1}{2}\lambda_R \phi_i^2}{\mu^2_R}
 \end{eqnarray}
 To account for the change from the initial temperature
to the final
 temperature ($T_i>T_c \; ; \; T_f<T_c$) we
parametrize\cite{note}
 \begin{eqnarray}
 m^2   & = & \mu_R(0)\left[\frac{T_i^2}{T_c^2} -1 \right]
\label{abovecrit} \\
 \mu_R & = & \mu_R(0)\left[1-\frac{T_f^2}{T_c^2} \right]
\label{belowcrit}
 \end{eqnarray}
 where the subscripts (R) stand for  renormalized
quantities, and
 $-\mu_R(0)$ is the renormalized zero temperature ``negative
mass squared''
 and $T^2_c = 24 \mu^2_R(0)/\lambda_R$.  Furthermore,
because $(k_{max}(t))^2
 \leq \mu^2_R$ and $T_i > T_c$, for the unstable modes $T_i
\gg (k_{max}(t))$
 and we can take the high temperature limit
$\mbox{coth}[\beta_i
 \omega_<(k)/2] \approx 2 T_i/\omega_<(k)$.  Finally the
effective
 equations of evolution for $t > 0$,
become, after using $\omega_<^2 = \mu_R^2(q^2+L^2)$ and
 keeping  only the unstable
 modes  as explained above ($q^2 <(q_{max}(\tau))^2$),

 \begin{eqnarray}
 \frac{d^2}{d\tau^2}\eta(\tau)-\eta(\tau)& + & \eta^3(\tau)+
\nonumber \\
                                         &   & g\eta(\tau)
\int_0^{(q_{max}(\tau))}q^2\frac{{\cal{U}}^+_q(\tau){\cal{U}}^-_q(\tau)}
  {q^2+L^2}dq = 0 \label{eqofmotunst}\\
  \left[\frac{d^2}{d\tau^2}+ q^2-
(q_{max}(t))^2\right]{\cal{U}}^{\pm}_q(\tau)
                                         & = & 0
\label{unstamod} \\
                       (q_{max}(\tau))^2 & = & 1-3\eta^2(\tau)
  \label{qmaxoft} \\
                                      g  & = & \frac{\sqrt{6\lambda_R}}
  {2\pi^2}\frac{T_i}{T_c \left[1-\frac{T^2_f}{T^2_c}\right]}
\label{coupling}
  \end{eqnarray}

For $T_i \geq T_c$ and $T_f \ll T_c$ the coupling (\ref{coupling}) is bound
within the range $g \approx 10^{-7}-10^{-8}$.
 The dependence of the coupling with the temperature
reflects the
 fact that at higher temperatures the fluctuations are
enhanced.

 From (\ref{eqofmotunst}) we see that the quantum
corrections act as a
 {\it positive dynamical renormalization} of the ``negative
mass'' term
 that drives  the rolling down dynamics. It is then clear,
that the
 quantum corrections tend to {\it slow down the evolution}.

In particular,
 if the initial value $\eta(0)$ is very small, the unstable
modes grow for
 a long time before $\eta(\tau)$ reaches the spinodal point
$\eta(\tau_s)
 = 1/\sqrt{3}$ at which point the instabilities shut off. If
this is the
 case, the quantum corrections introduce substantial
modifications to the
 classical equations of motion, thus becoming non-
perturbative. If
 $\eta(0)$ is closer to the classical spinodal point, the
unstable modes
 do not have time to grow dramatically and the quantum
corrections are
 perturbatively small.

 Thus we conclude that the initial conditions on the field
determine
 whether or not the quantum corrections are perturbatively
small.

 Figures (1,2) depict the solutions for the
classical
 (solid lines) and quantum (dashed lines) evolution.

 For the numerical integration we have
 chosen $L^2 =1$, the results are only weakly dependent on
$L$, and
 taken $g=10^{-7}$, we have varied the initial condition
$\eta(0)$ but
 used $\frac{d\eta(\tau)}{d\tau}|_{\tau=0} = 0$.

 We recall, from a previous discussion that $\eta(\tau)$ should be
 identified with the average of the field within a domain. We are
 considering the situation in which this average is very small, according
 to the usual slow-rollover hypothesis, and for  which the instabilities
 are stronger.

 In figure (1) we plot $\eta$ vs $\tau$ for $g=10^{-7}$,
 $\eta(0) = 2.3 \times 10^{-5}$ $\eta'(0) =0 \; ; \; L=1$.
The solid line is the classical evolution, the dashed line is the
evolution from the one-loop corrected equation of motion.
We begin to see that the
quantum  corrections become important at $t \approx 10/\mu_R$ and
slow down the dynamics. By the time that the {\it classical}
evolution  reaches the minimum of the classical potential at $\eta
=1$, the quantum evolution has just reached the classical spinodal
point $\eta = 1/\sqrt{3}$.
 The quantum correction becomes
large enough
 to change the sign of the ``mass term''\cite{boyveg}, the field
continues its
 evolution towards larger values, however, because the
velocity is
 different from zero.
As $\eta$ gets
 closer to the classical spinodal point, the unstability
shuts off  and the quantum correction
arising from the
 unstable modes become small. From the
spinodal point
 onwards, the field evolves towards the minimum and begins
to oscillate
 around it.
The quantum correction will be perturbatively
small, as all
 the instabilities had shut-off. Higher order corrections,
will introduce
 a damping term as quanta may decay into elementary
excitations of the true
 vacuum.

 Figure (2) shows a dramatic behavior for $\eta(0) =
2.258 \times
 10^{-5} \; ; \; \frac{d\eta (0)}{d\tau} = 0$ for the same values of the
parameters as for figure (1). The unstable
modes have
 enough time to grow so dramatically that the quantum
correction
  becomes extremely large
overwhelming the
 ``negative mass'' term near the origin.
 The dynamical time dependent potential, now becomes
 {\it a minimum} at the origin and the quantum evolution
begins to
 {\it oscillate} near $\eta = 0$. The contribution of the
unstable modes
 has become {\it non-perturbative}, and certainly our one-
loop
 approximation breaks down.

 As the initial value of the field gets closer to zero, the
unstable modes
 grow for a very long time. At this point, we realize,
however, that this
 picture cannot be complete. To see this more clearly, consider the case
 in which the
initial state
 or density matrix corresponds exactly to the symmetric
case. $\eta =0$ is
 necessarily, by symmetry, a fixed point of the equations of
motion.
 Beginning from the symmetric state, the field will {\it
always remain} at
 the origin and though there will be strong quantum and
thermal fluctuations,
 these are symmetric and will sample field configurations
with opposite
 values of the field with equal probability.

 In this situation, and according to the picture presented
above, one would
 then expect that the unstable modes will grow indefinetely
because the
 scalar field does not roll down and will never reach the
classical spinodal
 point thus shutting-off the instabilities. What is missing
in this picture
 and the resulting equations of motion is a self-consistent
treatment of the
 unstable fluctuations, which must necessarily go beyond one-
loop approximation.

\section{\bf Domain formation and growth:}

The instabilities correspond to the growth of long-wavelength
fluctuations and the formation of domains within which the
field is correlated. These domains will grow as long as the
instabilities persist. In order to understand the process of
domain growth when the average value of the field remains zero it
proves illuminating to understand the tree level correlations.

The relevant quantity of interest is the {\it equal time}
correlation function
\begin{eqnarray}
S(\vec{r};t) & = &
\langle\Phi(\vec{r},t)\Phi(\vec{0},t)\rangle
\label{equaltimecorr} \\
S(\vec{r};t) & = & \int \frac{d^3 k}{(2\pi)^3}e^{i\vec{k}
\cdot \vec{r}} S(\vec{k};t) \label{strucfac} \\
S(\vec{k};t) & = & \langle\Phi_{\vec{k}}(t)\Phi_{-
\vec{k}}(t)\rangle =
(-iG^{++}_{\vec{k}}(t;t))\label{strufack}
\end{eqnarray}
where we have performed the Fourier transform in the spatial
coordinates (there still is spatial translational and
rotational invariance). Notice that at equal times, all the
Green's functions are equal, and we may compute any of them.

Clearly in an equilibrium situation this equal time correlation function
will be time independent, and will only measure the
{\it static correlations}. In the present case, however, there is a non
trivial time evolution arising from the departure from equilibrium of the
initial state.

The unstable contribution to the Green function at equal times is
given by (\ref{earlytimegreen}), in this case for $\phi_i = 0$.

It is convenient to introduce the following
dimensionless quantities
\begin{equation}
\kappa = \frac{k}{m_f}  \; \; ; \; \;  L^2 = \frac{m_i^2}{m_f^2}=
\frac{\left[T^2_i-T^2_c\right]}{\left[T^2_c-T^2_f\right]}
\; \; ; \; \; \tau = m_f t  \; \; ; \; \;  \vec{x} = m_f \vec{r}
\label{dimensionless}
\end{equation}
Furthermore for the unstable modes $\vec{k}^2 < m^2_f$, and
for initial
temperatures  larger than the critical temperature
$T^2_c = 24 \mu^2 / \lambda$,  we can approximate $\coth[\beta_i
\omega_{<}(k)/2] \approx
2 T_i / \omega_{<}(k)$. Then, at tree-level, the
contribution of the
unstable modes to the subtracted structure factor
(\ref{strufack})
$S^{(0)}(k,t)-
S^{(0)}(k,0)=(1/m_f){\cal{S}}^{(0)}(\kappa,\tau)$ becomes
\begin{eqnarray}
{\cal{S}}^{(0)}(\kappa,\tau)   & = &
\left( \frac{24}{\lambda [1-\frac{T^2_f}{T^2_c}]} \right)^{\frac{1}{2}}
\left(\frac{T_i}{T_c} \right)
\frac{1}{2\omega^2_{\kappa}} \left( 1+
\frac{\omega^2_{\kappa}}{W^2_{\kappa}}
\right) \left[ \cosh(2W_{\kappa}\tau)-1 \right]
\label{strucuns} \\
              \omega^2_{\kappa}& = & \kappa^2+L^2 \label{smallomegak} \\
  W_{\kappa}                   & = & 1-\kappa^2 \label{bigomegak}
\end{eqnarray}

To obtain a better idea of the growth of correlations, it is
convenient to
introduce the scaled correlation function
\begin{equation}
{\cal{D}}(x,\tau) = \frac{\lambda}{6m^2_f}\int^{m_f}_0
\frac{k^2 dk}{2\pi^2}\frac{\sin(kr)}{(kr)}[S(k,t)-S(k,0)]
\label{integral}
\end{equation}
The reason for this is that the minimum of the tree level
potential occurs
at $\lambda \Phi^2 /6 m^2_f =1$, and the inflexion
(spinodal) point,
at $\lambda \Phi^2 /2 m^2_f =1$, so that ${\cal{D}}(0,\tau)$
measures the
excursion of the fluctuations to the spinodal point and
beyond as the correlations grow in time.

At large $\tau$ (large times),
the  product $\kappa^2 {\cal{S}}(\kappa,\tau)$ in
(\ref{integral}) has a very sharp peak at
$\kappa_s = 1/ \sqrt{\tau}$. In the region $x < \sqrt{\tau}$
the integral
may be done by the  saddle point approximation and we obtain for
$T_f/T_c \approx 0$ the large time behavior
\begin{eqnarray}
{\cal{D}}(x,\tau) & \approx & {\cal{D}}(0,\tau)
\exp[-\frac{x^2}{8\tau}]
\frac{\sin(x/ \sqrt{\tau})}{(x/ \sqrt{\tau})}
\label{strucfacxtau} \\
{\cal{D}}(0,\tau) & \approx & \left(\frac{\lambda}{12
\pi^3}\right)^
{\frac{1}{2}}\left(\frac{(\frac{T_i}{2 T_c})^3}{[
\frac{T^2_i}{T^2_c}-
1]}\right)\frac{\exp[2\tau]}{\tau^{\frac{3}{2}}}
\label{strucfactau}
\end{eqnarray}

 Restoring dimensions, and recalling that the zero
temperature correlation
 length is $\xi(0) = 1/\sqrt{2} \mu$,
 we find that for $T_f \approx 0$ the amplitude of the
fluctuation inside a
 ``domain'' $\langle \Phi^2(t)\rangle$, and the ``size'' of
a  domain  $\xi_D(t)$ grow as
 \begin{eqnarray}
 \langle \Phi^2(t)\rangle & \approx &  \frac{\exp[\sqrt{2}t/
\xi(0)]}
 {(\sqrt{2}t/ \xi(0))^{\frac{3}{2}}} \label{domainamplitude}
\\
 \xi_D(t)      & \approx & (8\sqrt{2})^{\frac{1}{2}}
 \xi(0)\sqrt{\frac{t}{\xi(0)}}
\label{domainsize}
 \end{eqnarray}

 An important time scale corresponds to the time $\tau_s$ at
which the  fluctuations
 of the field sample beyond the spinodal point. Roughly
speaking when this
 happens, the instabilities should shut-off as the mean
square root
 fluctuation of the field $\sqrt{\langle\Phi^2(t)\rangle}$
is now probing the stable  region.
 This will be seen explicitly below when we study the
evolution  non-perturbatively in the Hartree approximation and
the fluctuations are incorporated self-consistently in the evolution
equations.
 In zeroth order we estimate this time
from the condition
 $3{\cal{D}}(0,t) = 1$, we use
 $\lambda= 10^{-12} \; ; \;  T_i/T_c=2$,
 as representative parameters
 (this value
 of the initial temperature does not have  any particular physical
 meaning and was chosen only as  representative). We find
 \begin{equation}
 \tau_s \approx 10.15 \label{spinodaltime}
 \end{equation}
 or in units of the zero temperature correlation length
 \begin{equation}
 t \approx 14.2 \xi(0)
 \end{equation}
 for other values of the parameter $\tau_s$ is found from
the above condition
 on (\ref{strucfactau}).

Clearly any perturbative expansion will fail because the propagators
will contain unstable wavelengths, and the one loop term will grow
faster than the zero order, etc.\cite{boyshin}.
We now turn to a non-perturbative analysis
(for details see\cite{boyshin}).

As the correlations and fluctuations grow, field
configurations start sampling the stable region beyond the spinodal point.
This will result in  a slow down in the
growth of correlations, and eventually  the unstable growth
will shut-off.
When this happens, the state may be described by correlated domains
with equal
probabibility for both phases inside the domains. The
expectation value of
the field in this configuration will be zero, but inside
each domain, the
field will acquire a value very close to the value in
equilibrium at the
minimum of the {\it effective potential}. The size of the
domain in this
picture will depend on the time during which correlations
had grown enough so that
fluctuations start sampling beyond the spinodal point.

Since this physical picture may not be studied within
perturbation theory,
we now introduce a {\it non-perturbative} method based on a
self-consistent Hartree approximation, which is implemented as
follows:
in the initial Lagrangian write
\begin{equation}
\frac{\lambda}{4!}\Phi^4(\vec{r},t) =
\frac{\lambda}{4}\langle\Phi^2(\vec{r},t)\rangle
\Phi^2(\vec{r},t)+
\left(\frac{\lambda}{4!}\Phi^4(\vec{r},t)-
\frac{\lambda}{4}\langle\Phi^2(\vec{r},t)\rangle\Phi^2(\vec{
r},t)\right)
\label{hart}
\end{equation}
the first term is absorbed in a shift of the mass term
\[m^2(t) \rightarrow
m^2(t)+\frac{\lambda}{2}\langle\Phi^2(t)\rangle \]
(where we used spatial translational invariance). The second
term in
(\ref{hart}) is taken as an interaction with the term
$\langle\Phi^2(t)\rangle\Phi^2(\vec{r},t)$ as a ``mass
counterterm''.
The Hartree
approximation consists of  requiring that the one loop
correction to the two
point Green's functions must be cancelled by the ``mass
counterterm''. This
leads to the self consistent set of equations
\begin{equation}
\langle\Phi^2(t)\rangle  =  \int
\frac{d^3k}{(2\pi)^3}\left(-
iG_k^{<}(t,t)\right) =
\int \frac{d^3k}{(2\pi)^3} \frac{1}{2\omega_{<}(k)}
{\cal{U}}^+_k(t)
{\cal{U}}^-_k(t) \coth[\beta_i\omega_{<}(k)/2] \label{fi2}
\end{equation}
\begin{equation}
\left[\frac{d^2}{dt^2}+\vec{k}^2+m^2(t)+\frac{\lambda}{2}\langle
\Phi^2(t)\rangle\right]
{\cal{U}}^{\pm}_k=0 \label{hartree}
\end{equation}

Before proceeding any further, we must address the fact that
the composite
operator $\langle\Phi^2(\vec{r},t)\rangle$ needs one
subtraction and
multiplicative
renormalization. As usual the subtraction is absorbed in a
renormalization
of the bare mass, and the multiplicative renormalization
into a
renormalization of the coupling constant.

At this stage our justification
for using this approximation
is based on the fact that it provides a non-perturbative
framework to sum
an infinite series of Feynman diagrams of the cactus
 type\cite{dolan,chang}.

It is clear that for $t<0$ there is a self-consistent
solution to the
Hartree equations with equation (\ref{fi2}) and
\begin{eqnarray}
\langle\Phi^2(t)\rangle       & = &  \langle\Phi^2(0^-)\rangle \nonumber \\
{\cal{U}}^{\pm}_k             & = &  \exp[\mp i \omega_{<}(k)] \\
\omega^2_{<}(k)               & = &  \vec{k}^2+m^2_i+\frac{\lambda}{2}+
\langle\Phi^2(0^-)\rangle = \vec{k}^2+m^2_{i,R} \nonumber \\
\end{eqnarray}
where the composite operator has been absorbed in a renormalization of the
initial mass, which is now parametrized as
$m^2_{i,R}=\mu^2_R[(T^2_i/T^2_c)-1]$. For
$t>0$ we subtract the composite operator at $t=0$
absorbing the subtraction
into a renormalization of $m^2_f$ which we now parametrize
as $m^2_{f,R}=
\mu^2_R[1-(T^2_f/T^2_c)]$. We should point out that this
choice of
parametrization only represents a choice of the bare
parameters, which can
always be chosen to satisfy this condition. The logarithmic
multiplicative
divergence of the composite operator will be absorbed in a
coupling constant
renormalization consistent with the Hartree approximation\cite{chang,veg},
however, for the purpose of understanding the dynamics of growth of
instabilities associated with the long-wavelength fluctuations,
we will not need to specify this procedure. After
this subtraction
procedure, the Hartree equations read
\begin{equation}
[\langle\Phi^2(t)\rangle-\langle\Phi^2(0)\rangle]  =
\int \frac{d^3k}{(2\pi)^3} \frac{1}{2\omega_{<}(k)}
[{\cal{U}}^+_k(t)
{\cal{U}}^-_k(t)-1] \coth[\beta_i\omega_{<}(k)/2]
\label{subfi2}
\end{equation}
\begin{equation}
\left[\frac{d^2}{dt^2}+\vec{k}^2+m^2_R(t)+\frac{\lambda_R}{2}
\left(\langle\Phi^2(t)\rangle-\langle\Phi^2(0)\rangle\right)
\right]
{\cal{U}}^{\pm}_k(t)=0 \label{subhartree}
\end{equation}
\begin{equation}
m^2_R(t)= \mu^2_R \left[\frac{T^2_i}{T^2_c}-1\right]
\Theta(-t)
- \mu^2_R \left[1-\frac{T^2_f}{T^2_c}\right] \Theta(t)
\end{equation}
with $T_i > T_c$ and $T_f \ll T_c$.
With the self-consistent solution and boundary condition for
$t<0$
\begin{eqnarray}
[\langle\Phi^2(t<0)\rangle-\langle\Phi^2(0)\rangle]  & = & 0
\label{bcfi2} \\
{\cal{U}}^{\pm}_k(t<0)       & = & \exp[\mp i
\omega_{<}(k)t]
\label{bcmodes}\\
\omega_{<}(k)                & = & \sqrt{\vec{k}^2+m^2_{iR}}
\end{eqnarray}

This set of Hartree equations is extremely complicated
to be solved exactly.
However it has the correct physics in it. Consider the
equations for $t>0$,
at very early times, when (the renormalized) $\langle\Phi^2(t)\rangle-
\langle\Phi^2(0)\rangle \approx 0$
the mode functions are the same as in the zeroth order approximation,
and the unstable modes grow exponentially. By computing the expression
(\ref{subfi2}) self-consistently  with
these zero-order unstable modes, we see that the fluctuation
operator begins to grow exponentially.

As $(\langle\Phi^2(t)\rangle-\langle\Phi^2(0)\rangle)$ grows
larger,
its contribution to the Hartree equation tends to balance
the negative
mass term, thus weakening the unstabilities, so that only longer
wavelengths can become
unstable. Even for very weak coupling constants,
the exponentially
growing modes make the Hartree term in the equation of
motion for the mode
functions become large and compensate for the negative mass
term.
Thus when

\[\frac{\lambda_R}{2m^2_{f,R}}\left(\langle\Phi^2(t)\rangle-
\langle\Phi^2(0)\rangle\right)
\approx 1 \]
the instabilities
shut-off, this equality determines the ``spinodal time'' $t_s$.
The modes will still continue to grow further
after this point
because the time derivatives are fairly (exponentially)
large, but eventually
the growth will slow-down when fluctuations sample deep
inside the stable  region.

After the subtraction, and multiplicative renormalization
(absorbed in a
coupling constant renormalization), the composite operator
is finite. The
stable mode functions will make a {\it perturbative}
contribution to the
fluctuation which will be always bounded in time.  The most
important contribution will be that of the {\it unstable
modes}. These will grow
exponentially at early times and their effect will dominate
the dynamics of
growth and formation of correlated domains. The full set of
Hartree equations
is extremely difficult to solve, even numerically, so we
will restrict
ourselves to account {\it only} for the unstable modes. From
the above
discussion it should be clear that these are the only
relevant modes for the
dynamics of  formation and growth  of domains, whereas the
stable modes,  will
always contribute perturbatively for weak coupling after renormalization.

Introducing the dimensionless ratios (\ref{dimensionless})
in terms of
$m_{f,R}\; ; \; m_{i,R}$, (all momenta are now expressed in
units of
$m_{f,R}$), dividing (\ref{subhartree}) by $m_{f,R}^2$,
using the high temperature
approximation $\coth[\beta_i\omega_{<}(k)/2] \approx
2T_i/\omega_{<}(k)$
for the unstable modes, and expressing the critical
temperature as
$T^2_c=24 \mu_R/\lambda_R$, the set of Hartree equations
(\ref{subfi2},
\ref{subhartree}) become the following integro-differential
equation for
the mode functions for $t>0$
\begin{equation}
\left[\frac{d^2}{d\tau^2}+q^2-1+g\int^1_0 dp
\left\{\frac{p^2}{p^2+L^2_R}
[{\cal{U}}^+_p(t){\cal{U}}^-_p(t)-1]\right\}
\right]{\cal{U}}^{\pm}_q(t)=0
\label{finalhartree}
\end{equation}
with
\begin{eqnarray}
{\cal{U}}^{\pm}_q(t<0) & = & \exp[\mp i \omega_{<}(q)t]
\label{bounconhart} \\
\omega_{<}(q)          & = & \sqrt{q^2+L^2_R} \label{frequ}
\\
L^2_R                  & = & \frac{m^2_{i,R}}{m^2_{f,R}} =
\frac{[T_i^2-T_c^2]}{[T^2_c-T_f^2]}
\\
g                      & = & \frac{\sqrt{24\lambda_R}}{4\pi^2}
\frac{T_i}{[T^2_c-T^2_f]^{\frac{1}{2}}}
\label{effectivecoupling}
\end{eqnarray}
The effective coupling (\ref{effectivecoupling}) reflects
the enhancement of
quantum fluctuations by high temperature effects; for
$T_f/T_c \approx 0$,
and for couplings as weak as $\lambda_R \approx 10^{-12}$,
$g \approx 10^{-7} (T_i/T_c)$.

The equations (\ref{finalhartree}) may now be integrated
numerically for the
mode functions; once we find these, we can then compute the
contribution of the unstable modes
 to the subtracted
correlation function equivalent to  (\ref{integral})
\begin{eqnarray}
{\cal{D}}^{(HF)}(x,\tau)    & = &
\frac{\lambda_R}{6 m_{f,R}^2} \left[\langle \Phi(\vec{r},t)
\Phi(\vec{0},t)\rangle-
\langle\Phi(\vec{r},0)\Phi(\vec{0},0)\rangle\right]
\label{hartreecorr1} \\
3{\cal{D}}^{(HF)}(x,\tau)   & = & g\int_0^1dp
\left(\frac{p^2}{p^2+L^2_R}
\right)\frac{\sin(px)}{(px)}\left[{\cal{U}}^+_p(t){\cal{U}}^
-_p(t)-1\right]
\label{hartreecorr2}
\end{eqnarray}
In figure (3) we show

\[ \frac{\lambda_R}{2m_{f,R}^2}(\langle\Phi^2(\tau)\rangle -
\langle\Phi^2(0)\rangle)=
3({\cal{D}}^{HF}(0,\tau)- {\cal{D}}^{HF}(0,0)) \]
(solid line) and
also for comparison, its zeroth-order counterpart
$3({\cal{D}}^{(0)}(0,\tau)-{\cal{D}}^{(0)}(0,0))$ (dashed
line)
for $\lambda_R = 10^{-12}\; , \; T_i/T_c=2$.
(this value of the initial temperature does not have any
particular physical significance and was chosen as a
representative).  We clearly see what we expected;
 whereas the zeroth order correlation grows indefinitely,
the Hartree
correlation function is bounded in time and oscillatory. At
$\tau \approx
10.52$ ,  $3({\cal{D}}^{(HF)}(0,\tau)-{\cal{D}}^{(HF)}(0,\tau))
= 1$,
fluctuations are sampling field configurations near the
classical spinodal, fluctuations
 continue to grow, however,  because the
derivatives are still fairly large. However, after this time, the
modes  begin  to probe the stable region in which there is no
exponential growth. At this point
$\frac{\lambda_R}{2m_{f,R}^2}(\langle\Phi^2(\tau)\rangle-
\Phi^2(0)\rangle)$,
becomes small again because of the small coupling $g \approx
10^{-7}$, and the correction term becomes small.  When
it becomes
smaller than one, the instabilities set in again, modes
begin to grow and the process repeats.
This gives rise to an oscillatory behavior around
$\frac{\lambda_R}{2m^2_{f,R}}(\langle\Phi^2(\tau)\rangle-
\Phi^2(0)\rangle=1$ as shown in figure (3).
In figure (4), we show the structure factors as a
function of
$x$ for $\tau =  10$, both for zero-order
(tree
level) ${\cal{D}}^{(0)}$ (dashed lines) and Hartree
${\cal{D}}^{(HF)}$ (solid lines) for the same value of the parameters as
for figure (3).
These correlation functions clearly show the
growth in amplitude
 and that the size of the region in which
the fields are
correlated increases with time. Clearly this region may be
interpreted as
a ``domain'', inside which the fields have strong
correlations, and outside
which the fields are uncorrelated.

We see that up to the spinodal time $\tau_s \approx 10.52$
at which
 $\frac{\lambda_R}
{2m_{f,R}}(\langle\Phi^2(\tau_s)\rangle-
\Phi^2(0)\rangle)=1$, the zeroth order
correlation
$3{\cal{D}}^{(0)}(0,\tau)$ is very close to the Hartree
result. In fact at
$\tau_s$, the difference is less than $15\%$.  For
these values of the coupling and initial temperature, the zeroth
order correlation function leads to $t_s \approx 10.15$, and we may
use the zeroth order correlations to provide an analytic estimate for
$t_s$, as well as  the form of the correlation functions and the
size of the  domains.
The fact that the zeroth-order
correlation remains very close to the Hartree-corrected
correlations up to
times comparable to the spinodal is a consequence of
the very small coupling.
The
stronger coupling makes
the growth of domains much faster and the departure from
tree-level
correlations more dramatic\cite{boyshin}. For
strong couplings
domains will form very rapidly and only grow to sizes of the
order of the
zero temperature correlation length. The phase transition
will occur very
rapidly, and  our initial assumption of a rapid
supercooling will be unjustified.
 This situation for strong couplings, of domains
forming very
rapidly to sizes of the order of the zero temperature
correlation length
is the picture presented by Mazenko and collaborators\cite{mazenko}.
However, for very
weak couplings (consistent with the bounds from density
fluctuations), our
results indicate that the phase transition will proceed very
slowly, domains
will grow for a long time and become fairly large,
with a typical size several times the zero
temperature correlation length. In a sense, this is a self
consistent check
of our initial assumptions on a rapid supercooling in the
case of weak couplings.

As we argued above, for very weak coupling
we may use the tree level result to give an
approximate bound to the correlation functions up to times
close to the
spinodal time using the result given by equation (\ref{strucfactau}),
for $T_f \approx 0$.
Thus, we conclude that for large times, and very weakly
coupled theories ($\lambda_R \leq 10^{-12}$) and for initial temperatures
of the order of the critical temperature,
the size of the domains $\xi_D(t)$ will grow typically in time
as

\begin{equation}
 \xi_D(t)       \approx  (8\sqrt{2})^{\frac{1}{2}}
 \xi(0)\sqrt{\frac{t}{\xi(0)}}
\label{sizedomain}
 \end{equation}

\noindent with $\xi(0)$ the zero temperature correlation length. The
maximum size of
a domain is approximately determined by the time at which
fluctuations begin
probing the stable region, this is the spinodal time $t_s$ and the
maximum size of the domains is approximately $\xi_D(t_s)$.

An estimate for the spinodal time, is obtained from equation
(\ref{strucfactau}) by the condition $3{\cal{D}}(\tau_s)=1$.
 For weakly
coupled theories and $T_f \approx 0$, we obtain
\begin{equation}
\tau_s = \frac{t_s}{\sqrt{2}\xi(0)} \approx -\ln\left[
\left(\frac{3\lambda}{4\pi^3}\right)^
{\frac{1}{2}}\left(\frac{(\frac{T_i}{2 T_c})^3}{[
\frac{T^2_i}{T^2_c}-
1]}\right)\right]
\end{equation}

\section{FRW Cosmologies:}

We now consider the  case of scalar field in an homogeneous, isotropic
and spatially flat FRW cosmology described by the metric
\be
ds^2 = dt^2-a^2(t)d\vec{x}^2
\ee
 The action and Lagrangian density are given by
\ba
S         & =  & \int d^4x {\cal{L}} \label{action} \\
{\cal{L}} & =  & a^3(t)\left[\frac{1}{2}\dot{\Phi}^2(\vec{x},t)-
\frac{1}{2}
\frac{(\vec{\nabla}\Phi(\vec{x},t))^2}{a(t)^2}-V(\Phi(\vec{x},t))\right]
 \label{lagrangian} \\
V(\Phi)   & =  & \frac{1}{2}[m^2+ \xi {\cal{R}}] \Phi^2(\vec{x},t)+
\frac{\lambda}{4!}\Phi^4(\vec{x},t) \label{potential} \\
{\cal{R}}    & =  & 6\left(\frac{\ddot{a}}{a}+\frac{\dot{a}^2}{a^2}\right)
\ea
with ${\cal{R}}$ the Ricci scalar. We have introduced the coupling
to the Ricci scalar as it will be induced by renormalization (see
below).

We can either follow the same method outlined above in terms of a path
integral in the complex time plane, or alternatively one could evolve
the density matrix in time by solving explicitly the Liouville equation,
these two methods are equivalent.

We have followed\cite{boyveghol} the second alternative in one-loop and
 Hartree approximation (thus
providing another independent manner of tackling the problem).

The initial condition is assumed to correspond to a local equilibrium
thermal  ensemble  of the adiabatic modes at an early time
$t_o \rightarrow -\infty$
 at a temperature $T_o = 1 / \beta_o$. In the case of a de Sitter
background, this initial condition corresponds to a thermal distribution
of the Bunch-Davies modes, and for $T_o \rightarrow 0$ the density
matrix describes a pure state, the Bunch-Davies vacuum.

Any description of quantum statistical mechanics needs to specify an
initial density matrix whose subsequent time evolution is determined
by the Hamiltonian. The assumption of local thermal equilibrium at an
early time is somewhat arbitrary. Whether or not it corresponds to
a physically realistic situation can only be answered within the realm
of a theory that incorporates gravity, particle physics and
statistical mechanics  as  dynamical ingredients. The
choice of an initial condition for a density matrix will pervade any
out of equilibrium
dynamical calculation. If the system reaches thermal
equilibrium after a long time, the details on the initial conditions
may ultimately be irrelevant, but this requires a deeper understanding.

Our goal is to obtain the evolution equation for the order parameter,
as well as the fluctuations. The order parameter is the expectation value
(in the time dependent density matrix) of the volume average of the
scalar field
\be
\phi(t) = \frac{1}{\Omega}\int d^3x \langle \Phi(\vec{x},t) \rangle =
\frac{1}{\Omega}\int d^3x  Tr\hat{\rho}(t)\Phi(\vec{x}) =
\frac{1}{\Omega}\int d^3x Tr \hat{\rho}(t_o)U^{-1}(t,t_o)\Phi(\vec{x},t_o)
U(t,t_o)
\label{orderparameter}
\ee
where $\Omega$ is the comoving volume, and the scale factors cancel
between the numerator (in the integral) and the denominator.
$U(t,t_o)$ is the time evolution operator with boundary condition
$U(t_o,t_o)=1$, $t_o$ is the time at which the initial ensemble is
specified.

To one-loop order we find the effective evolution equations for the
order parameter\cite{boyveghol}
\begin{equation}
\ddot{\phi}+3 \frac{\dot{a}}{a}\dot{\phi}+V'(\phi)+\lambda\phi
\frac{\hbar}{2}\int \frac{d^3k}{(2\pi)^3}
 \frac{\left[{\cal{U}}^2_{1k}(t)+
{\cal{U}}^2_{2k}(t)\right]}{2a^3(t)W_k(t_o)}\coth\left[\beta_o
\hbar W_k(t_o)/2
\right] = 0 \label{finaleqofmot}
\end{equation}
where the mode functions ${\cal{U}}_{\alpha,k}(t) \; ; \; \alpha=
1,2$ are real and satisfy
\begin{eqnarray}
& & \ddot{\cal{U}}_{\alpha k}-\frac{3}{2}\left(\frac{\ddot{a}}{a}+
\frac{1}{2}
\frac{\dot{a}^2}{a^2}\right){\cal{U}}_{\alpha k}+\left(
\frac{\vec{k}^2}{a^2(t)}+V''(\phi_{cl}(t)\right)
{\cal{U}}_{\alpha k}
=0 \label{diffeqU} \\
& & {\cal{U}}_{1k}(t_o) = 1 \; \; ; \; \; {\cal{U}}_{2k}(t_o) = 0
\label{boundconU} \\
& & \dot{{\cal{U}}}_{1k}(t_o) = \frac{3}{2}\frac{\dot{a}(t_o)}{a(t_o)}
\; \; ; \; \; \dot{{\cal{U}}}_{2k}(t_o) = W_k(t_o) =
\left[\frac{\vec{k}^2}{a^2(t_o)}+V''(\phi_{cl}(t_o))\right]^{\frac{1}{2}}
\label{boundconUdot}
\end{eqnarray}
\noindent where $\phi_{cl}(t)$ is the solution to the {\it classical}
equations of motion.

We can see clearly that the time dependence of the mode functions is very
important. If the scale factor varies rapidly with time as for example
in a radiation dominated or de Sitter epoch,
 at no time will the notion of an effective potential be valid
for the time evolution of the order parameter.

The method lends itself to a self-consistent non-perturbative Hartree
approximation.
Invoking a Hartree factorization as in a previous section, we also
derived the equations of motion. Introducing the fluctuation
 $\eta(\vec{x},t) = \Phi(\vec{x},t)-\phi(t)$ and the
 Hartree ``frequencies''
\begin{equation}
{\cal{V}}^{(2)}(\phi) = V''(\phi)+\frac{\lambda}{2} \langle \eta^2 \rangle
\end{equation}
we find
\begin{eqnarray}
& & \ddot{\phi}+3 \frac{\dot{a}}{a}\dot{\phi}+V'(\phi)+
\lambda \phi \frac{\hbar}{2}  \int \frac{d^3k}{(2\pi)^3}
 \frac{\left[{\cal{U}}^2_{1k}(t)+
{\cal{U}}^2_{2k}(t)\right]}{2a^3(t){\cal{W}}_k(t_o)}
\coth\left[\beta_o \hbar {\cal{W}}_k(t_o)/2
\right] = 0 \label{harteqofmot} \\
& & \left[\frac{d^2}{dt^2}-\frac{3}{2}\left(\frac{\ddot{a}}{a}+
\frac{1}{2}
\frac{\dot{a}^2}{a^2}\right)+
\frac{\vec{k}^2}{a^2(t)}+V''(\phi(t)) + \right. \nonumber \\
& & \left. \lambda \frac{\hbar}{2}
 \int \frac{d^3k}{(2\pi)^3}
\frac{\left[{\cal{U}}^2_{1k}(t)+
{\cal{U}}^2_{2k}(t)\right]}{2a^3(t){\cal{W}}_k(t_o)}
\coth\left[\beta_o \hbar {\cal{W}}_k(t_o)/2
\right] \right]
{\cal{U}}_{\alpha k}
=0 \label{newdiffeqU} \\
& & {\cal{U}}_{1k}(t_o) = 1 \; \; ; \; \; {\cal{U}}_{2k}(t_o) = 0
\label{newboundconU} \\
& & \dot{{\cal{U}}}_{1k}(t_o) = \frac{3}{2}\frac{\dot{a}(t_o)}{a(t_o)}
\; \; ; \; \; \dot{{\cal{U}}}_{2k}(t_o) = {\cal{W}}_k(t_o) =
\left[\frac{\vec{k}^2}{a^2(t_o)}+ {\cal{V}}^{(2)}(\phi(t_o))
\right]^{\frac{1}{2}}
\label{newboundconUdot}
\end{eqnarray}

The one-loop and Hartree equations present new features not present in
our previous analysis in Minkowski space:
 first the ``friction'' term arising
from the expansion, and secondly the fact that physical wave-vectors
are red-shifted. The red-shift will tend to enhance the instabilities
as more wavelengths become unstable, but the presence of the horizon will
ultimately constrain the final size of the correlated regions.

We are currently studying the numerical evolutions of these equations
and expect to report soon on details of the dynamics of the phase
 transition in these cosmologies\cite{boyveghol}.

\section{\bf Renormalization Aspects}

To understand the renormalization procedure, it proves convenient
 to introduce the
complex combination of mode functions
\ba
{\cal{U}}^{\pm}_k(t)       & = & {\cal{U}}_{1k}\mp i {\cal{U}}_{2k}
\label{ucomplex} \\
{\cal{U}}^{\pm}_k(t_o)     & = & 1 \label{boundcomplex} \\
{\dot{\cal{U}}}^{\pm}_k(t) & = & \frac{3\dot{a}(t_o)}{2a(t_o)}
\mp i {\cal{W}}_k(t_o) \label{dotboundcomp}
\ea
in the case of the one-loop approximation ${\cal{W}}_k(\phi)$
 must be replaced
by $W_k(\phi)$.
We will now analyze the renormalization aspects for the Hartree
 approximation, the one-loop case may be obtained easily from this more
general case.
We need to understand the divergences in the integral
\be
I =  \int \frac{d^3k}{(2\pi)^3}
\frac{{\cal{U}}^+_{k}(t)
{\cal{U}}^-_{k}(t)}{2a^3(t){\cal{W}}_k(t_o)} \label{divint}
\ee
The divergences in this integral will be determined from the large-k
behavior of the mode functions that are solutions to the differential
equation (\ref{newdiffeqU}) with the boundary conditions
(\ref{boundcomplex}, \ref{dotboundcomp}). The large-k behavior of these
functions may be obtained in a WKB approximation by introducing the
WKB function
\ba
{\cal{D}}_k(t)   & = & \exp\left[\int^t_{t_o} R(t')dt'\right] \label{wkb} \\
{\cal{D}}_k(t_o) & = & 1 \label{wkbboundcon}
\ea
satisfying the differential equation
\be
\left[\frac{d^2}{dt^2}-\frac{3}{2}\left(\frac{\ddot{a}}{a}+
\frac{1}{2}
\frac{\dot{a}^2}{a^2}\right)+
\frac{\vec{k}^2}{a^2(t)}+V''(\phi(t)) + \frac{\lambda}{2} \langle \eta^2
(\vec{x},t)\rangle
\right]
{\cal{D}}_{ k}
=0 \label{wkbequ}
\ee
with $\langle eta^2 (\vec{x},t) \rangle$
 being the self-consistent integral in the Hartree
equation (\ref{newdiffeqU}) (the one-loop approximation is obtained by
setting this term to zero in the above equation). The mode functions of
 interest
${\cal{U}}^{pm}$ are obtained as linear combinations
of the WKB function and its complex conjugate. The coefficients
to be determined from the boundary condition at $t_o$. The function
$R(t)$ obeys a Riccati  (WKB) equation
\be
\dot{R}+R^2-\frac{3}{2}\left(\frac{\ddot{a}}{a}+
\frac{1}{2}
\frac{\dot{a}^2}{a^2}\right)+
\frac{\vec{k}^2}{a^2(t)}+V''(\phi(t)) +
\frac{\lambda}{2} \langle \eta^2
(\vec{x},t)\rangle = 0
\label{wkbriccati}
\ee
We propose a solution to this equation of the form
\be
R= \frac{-i k}{a(t)}+R_o(t)-\frac{i R_1(t)}{k}+\frac{R_2(t)}{k^2}+\cdots
\label{wkbseries}
\ee
and find the time dependent coefficient by comparing powers of k. We
find
\ba
R_o(t) & = & \frac{\dot{a}(t)}{2a(t)} \label{Ro} \\
R_1(t) & = & \frac{a(t)}{2}\left[-\frac{{\cal{R}}}{6}+V''(\phi(t))+
\frac{\lambda}{2} \langle \eta^2
(\vec{x},t)\rangle
 \right] \label{R1} \\
R_2(t) & = & -\frac{1}{2}\frac{d}{dt}\left[a(t)R_1(t)\right] \label{R2}
\ea

Finally we obtain

\be
I =  \frac{1}{8\pi^2} \frac{\Lambda^2}{a^2(t)}+
  \frac{1}{8\pi^2}
\ln \left(\frac{\Lambda}{K}\right)
 \left[\frac{\dot{a}^2(t_o)}{a^2(t)}-\left(-\frac{{\cal{R}}}{6}+
V''(\phi(t))+ \frac{\lambda}{2} \langle \eta^2
(\vec{x},t)\rangle
\right)\right]+ \mbox{ finite } \label{divergences}
\ee
where we have introduced a renormalization point $K$, and the finite part
depends on time, temperature and $K$.
For renormalization in the one-loop approximation,
 $\langle \eta^2 \rangle$  does not
appear in the logarithmic divergent term in (\ref{divergences}) (as it
does not appear in the differential equation for the mode functions up
to one loop).
There are several physically important features of the divergent structure
obtained above. The first term (quadratically divergent) reflects
 the fact that the physical momentum cut-off is being red-shifted by the
expansion. This term will not appear in dimensional regularization.

Secondly, the logarithmic divergence contains a term that reflects the
initial condition (the derivative of the expansion factor at the initial
time $t_o$). The initial condition breaks any remnant symmetry (for example
in de Sitter space there is still invariance under the de Sitter group,
but this is also broken by the initial condition at an arbitrary time
$t_o$).
Thus this term is not forbidden, and its appearance does not come as a
surprise.

The renormalization conditions in the Hartree approximation are
obtained by requiring that the equation for the mode functions be
finite\cite{boyveghol}. Thus we obtain
\ba
& & m^2_B +\frac{\lambda_B\hbar}{16\pi^2}\frac{\Lambda^2}{a^2(t)}+
\frac{\lambda_B\hbar}{16\pi^2}\ln \left(\frac{\Lambda}{K}\right)
 \frac{\dot{a}^2(t_o)}{a^2(t)}  =
m^2_R\left[1+\frac{\lambda_B\hbar}{16\pi^2}
\ln \left(\frac{\Lambda}{K}\right)\right] \label{massren} \\
& & \lambda_B                       =
\lambda_R \left[1+\frac{\lambda_B\hbar}{16\pi^2}
\ln \left(\frac{\Lambda}{K}\right)\right] \label{lambdaren} \\
& & \xi_B                         =
\xi_R + \frac{\lambda_B\hbar}{16\pi^2}
\ln \left(\frac{\Lambda}{K}\right) \left(\xi_R-\frac{1}{6}\right)
\label{xiren}
\ea

Notice that the conformal coupling $\xi = 1/6$ is a {\it fixed point}
under renormalization.

\section{\bf High Temperature Limit:}

One of the payoffs of understanding the large-k behavior of the mode
functions (as obtained in the previous section via the WKB method) is
that it permits a straightforward evaluation of the high temperature
limit.
The finite temperature contribution is determined by the integral
\be
I_{\beta_o} =
\int \frac{d^3k}{(2\pi)^3}
 \frac{\left[{\cal{U}}^2_{1k}(t)+
{\cal{U}}^2_{2k}(t)\right]}{a^3(t)W_k(t_o)}
\frac{1}{e^{\beta_o
\hbar {\cal{W}}_k(t_o)}-1} \label{tempint}
\ee
For large temperature, only momenta $k \geq T_o$ contribute. Thus the
leading contribution is determined by the first term of the function
$R(t)$ of the previous section. After some straightforward algebra
we find
\be
I_{\beta_o} = \frac{1}{12}
\left[ \frac{k_B T_o a(t_o)}{\hbar a(t)}\right]^2
+{\cal{O}}(1/T_o) + \cdots \label{highT}
\ee

Thus we see that the leading high temperature behavior reflects the
physical
red-shift, in the cosmological background
and it results in an effective time dependent temperature

\[ T_{eff}(t) = T_o \left[\frac{a(t_o)}{a(t)}\right] \]

To this leading order, the expression obtained for the time dependent
effective temperature corresponds to what would be obtained for an
{\it adiabatic} (isentropic) expansion for blackbody-type radiation (massless
relativistic particles) in the cosmological background.

This behavior, however, only appears at {\it leading}
order in the high temperature expansion. Higher order terms do not seem
to have this form\cite{boyveghol} as both the coupling to the curvature and
the mass term distort the spectrum from the blackbody form.

\section{\bf Conclusions:}

We provided a dynamical picture of the time evolution of a weakly
coupled inflaton scalar field theory undergoing a typical second order
phase transition in Minkowski space-time. The shortcomings of the
usual approach based on the equilibrium effective potential
 and the necessity for
a description out of equilibrium were pointed out.
The non-equilibrium evolution equations for the order parameter were
derived to one-loop and integrated for ``slow-rollover'' initial
conditions. We pointed out that the instabilities responsible for the
onset of the phase transition and the process of domain formation,
 that is the growth of long-wavelength fluctuations, introduced dramatic
corrections to the ``slow-rollover'' picture. The net effect of these
instabilities is to {\it slow even further} the evolution of the order
parameter (average of the scalar field) and for ``slow rollover'' initial
conditions (the scalar field very near the false vacuum initially) the
dynamics becomes {\it non-perturbative}.

We introduced a self-consistent approach to study the dynamics out of
equilibrium in this situation. It is found that for very weakly coupled
theories domains growth to large sizes typically several times the
zero temperature correlation length and that the growth obeys a scaling
law at intermediate times.

In strongly coupled theories the phase transition will occur much faster
and domains will only grow to sizes of the order of the zero temperature
correlation length.

The one-loop equations of evolution for the scalar order parameter were
obtained in spatially flat FRW cosmologies. These equations reveal that
at no times is the approximation of an effective potential valid.
We have also obtained the evolution equations in a non-perturbative
self-consistent Hartree approximation. These equations present new
features that illuminate the
fact that the process of phase separation in cosmological settings is
more subtle.

 The physical wave vectors undergo a red-shift and more
wavelengths become unstable, thus enhancing the instability. On the
other hand the presence of the horizon and the ``friction'' term in
the equations of motion, will prevent domains from growing bigger
 than the
horizon size. Eventually there is a competition of time scales that
must be understood more deeply to obtain any meaningful conclusion about
formation and growth of domains in FRW cosmologies. Work on these issues
is in progress\cite{boyveghol}.

The renormalization aspects were studied and we point out that in
a time dependent background, divergences appear that depend on the
initial conditions. We also studied the leading behavior in the
 high temperature limit showing to this order that the temperature is
red-shifted as in adiabatic expansion.

This formalism could also be applied to chaotic initial
conditions\cite{linde}

{\bf Acknowledgments:}

D. B. would like to thank, D-S. Lee,  and A. Singh for
fruitful discussions and remarks, LPTHE at the Universite de
Pierre et Marie Curie for hospitality during part of this work,
and the N.S.F. for support through Grant: PHY-9302534 and for a
Binational Collaboration supported by N.S.F. through Grant: INT-9216755.


\begin{thebibliography}{99}
\bibitem{note} This may be done formally by generalizing the method proposed
by S. Weinberg\cite{sweinberg} and introduce a time dependent counterterm
in the original Hamiltonian of the form $\frac{\lambda}{24}\left(T_i^2
\theta(-t)+T_f^2 \theta(t)\right)$ and cancelling this counterterm against
the one loop contribution for $t<0$.
\end{thebibliography}
\end{document}